\documentclass[prb,twocolumn,amsmath,amssymb,superscriptaddress]{revtex4}
\usepackage[english]{babel}
\usepackage{amsmath}
\usepackage{amsfonts}
\usepackage{graphicx}
\usepackage{color}

\newcommand{\beq}{\begin{equation}} 
\newcommand{\eeq}{\end{equation}}

\newcommand{\nn}{\nonumber}

\newcommand{\bs}{\boldsymbol}

\begin{document}
\title {Dynamics of a Majorana  trijunction in a microwave cavity }

\author{\it Mircea Trif}
\affiliation{International Research Centre MagTop, Institute of Physics, Polish Academy of Sciences,
Aleja Lotnikow 32/46, PL-02668 Warsaw, Poland}

\author{\it Pascal Simon}
\email{pascal.simon@u-psud.fr}
\affiliation{Laboratoire de Physique des Solides, CNRS, Univ. Paris-Sud,
Universit\'e Paris Saclay, 91405 Orsay cedex, France}
\date{\today}

\begin{abstract}

A trijunction made of three topological semiconducting wires, each supporting a Majorana bound state  at its two extremities,  appears as one of the simplest geometry in order to perform braiding of Majorana fermions. By embedding the trijunction into a microwave cavity allows to study the intricate dynamics of the low-energy  Majorana bound states (MBSs) coupled to the cavity electric field under a braiding operation.  
Extending a previous work (Phys. Rev. Lett. 2019, 122, 236803),  the full time evolution of the density matrix of the low-energy states, including various relaxation channels, is computed both in the  adiabatic regime, as well as within the Floquet formalism in the case of periodic driving. It turns out that in the stationary state the observables of the system  depend on both the parity of the ground state and on the non-Abelian Berry phase acquired during braiding.
 The average photon number and the second order photon coherence function $g^{(2)}(0)$ are explicitly evaluated and reveal the accumulated non-Abelian Berry phase during the braiding process.   
\end{abstract}

\maketitle

\section{Introduction} 

Majorana quasiparticles  occur as zero energy excitations in topological superconductors.\cite{kitaev2001unpaired} They are presently the subject of an intense study (see {\it e.g.} Refs. [\onlinecite{alicea2012new,franz2015review,Aguado2017}] for reviews).  Many platforms have been proposed theoretically and are currently experimentally studied. This includes semiconducting wires proximitized by a s-wave superconductor, 
\cite{mourik2012signatures,xu2012,Das2012,albrecht2016exponential,Deng2016,Kouwenhoven2018,Lutchyn2018} arrays of magnetic atoms on top of a superconducting substrate,\cite{Yazdani2014} iron-based superconductors,\cite{Wang2018} or in proximitized second order topological insulators.\cite{Yazdani2019}
Localized Majorana bound states (MBSs) manifest as a degenerate ground state in the spectrum of a topological superconductor. They are protected in energy by a gap and spatially separated  by some macroscopic distance,
which therefore renders them more immune  against local perturbations.
For example, a topological semiconducting wire hosts two MBSs, one at each extremity of the wire.
The 2-fold degeneracy of the ground state encodes the fermion parity of the system i.e. the fermion occupation of the ground state. Such ground state degeneracy confer to
to MBSs non-Abelian statistics  under braiding operations.\cite{Kitaev2003,Alicea2011,Sau2011,vanHeck2012}
All these properties therefore make MBSs ideal candidates to realize topologically protected qubits.\cite{SarmaRMP2008,dassarma-majarona-review,aasen2016-qc}

Here, we consider the simplest geometry that realize braiding of MBSs:\cite{KarzigPRX2016} it consists of a trijunction (also called Y-junction)  as sketched in Fig. 1 made of three topological semiconducting wires embedded in a microwave cavity. Assuming all wires are in a topological phase, the maximum number of MBSs is six. However, due to the central overlap between the three wires, the low-energy degrees of freedom reduce to four MBSs (denoted  $\gamma_j,\,j=0-3$ in Fig. 1).\cite{vanHeck2012}
These four MBSs {\it a priori} encode two fermionic states. For a given fixed parity of the system, the number of low-energy states is again halved and  the low-energy sector of this system can therefore be mapped to a spin 1/2 in an effective dynamical magnetic field which incorporates both the knobs of the wires but also  the internal cavity field.\cite{Trif2019}

We have shown in a previous paper that both the parity of the ground state and the Berry phase associated with the braiding statistics  are imprinted into the cavity electric field via a frequency shift.\cite{Trif2019}  By probing the microwave cavity by standard reflectometry, such shift can be measured by conventional dispersive readouts techniques. The present  Berry phase coupling mechanism, which is due to the interplay of dynamics during the braiding protocol and the non-locality of the photonic field, works even when the lowest energy subspace spanned by the Majorana fermions is degenerate at all times. These effects are purely dynamical and do not have any static analogue.

In this paper, we go one step further and  exploit the mapping between the low-energy Hamiltonian describing the set-up  of Fig. 1 to a spin 1/2 in a time dependent magnetic field to calculate and analyze the full dynamics of the low-energy density matrix of the Y-junction during a braiding protocol in the presence of dissipation mechanisms. Using the Floquet-Markov approximation to describe the spin precession dynamics,
we obtain a time-dependent master equation for the  reduced density matrix which allows to evaluate, in a realistic fashion, both the dynamics of the MBSs in the cavity and that of the cavity field itself.  Two different approaches to compute the dynamics of the low-energy density matrix during a braiding bath in parameter space are used: either the  adiabatic limit is considered, or the Floquet formalism in the case of periodic driving. Within both approaches and in the stationary limit, we show that  photonic observables such as the photon number and  the coherence function depend on the non-Abelian Berry phase and on parity of the system.

The plan of the paper is as follows: In Sec.~\ref{sec1} we introduce the effective model Hamiltonian describing the dynamics of MBSs in a trijunction geometry coupled to a cavity field. Here, we also show how this model maps to that of a (slowly) precessing spin and eventually to a model similar to Rabi Hamiltonian (we coin this model Majorana-Rabi model). In Sec.~\ref{sec2} we discuss the dynamics of the combined system during the braiding and in the presence of dissipation. Specifically, we first discuss the adiabatic Master equation and calculate the first non-adiabatic corrections to the density matrix that allow us to extract various observables, such as the average number of photons in the cavity. We show that in leading order in the dynamics, the observables are dressed with the parity-dependent Berry phase contributions. Second, we address the braiding dynamics within a Floquet formalism in the presence of dissipation, and evaluate the average photon number in the stationary limit, and compare the two  approaches. In Sec.~\ref{discussion}, we discuss  whether the cavity is able to discriminate MBSs from quasi-zero Andreev-states (which are giving similar signatures in transport experiments) during braiding operations and also provide some experimental considerations on the  braiding frequency and other energy scales. Finally, in Sec.~\ref{conclusion} we end up with some conclusions and outlook. 
\begin{figure}[t] 
\includegraphics[width=\linewidth]{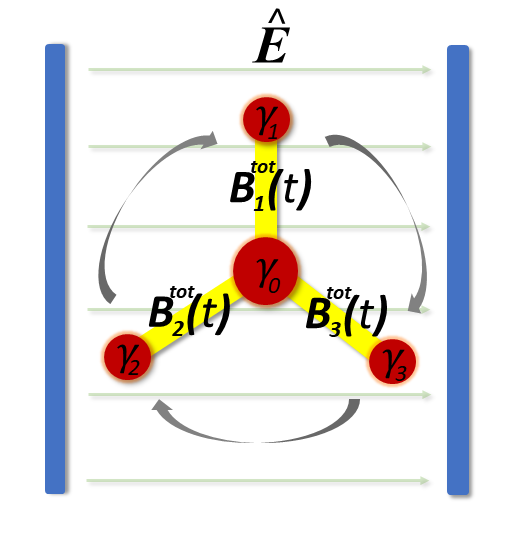}
\caption{ The Y junction consists of end MBSs $\gamma_{j}$, $j=1,2,3$,   and the middle $\gamma_0$ MBS, along with the couplings $B_{j}^{\rm tot}(t)=B_j(t)+g_j(a^\dagger+a)$ in the presence of the cavity field $\vec{E}$. }
\label{fig1} 
\end{figure}

\section{System and Model Hamiltonian}
\label{sec1}

Our starting point is a Majorana Y junction made of three topological semiconducting wires. This constitutes a minimal setup in order to perform non-Abelian braiding of Majorana bound states.\cite{Alicea2011,vanHeck2012,KnappPRX2016} We assume that  this device is embedded into  a microwave cavity, as depicted in Fig.~\ref{fig1} and couples to the cavity electromagnetic field. We assume that all three semiconducting wires consist of topological superconductors deep in the topological phase. They  thus all host  MBSs localized at their edges and protected in energy by a proximity induced superconducting gap $\Delta_{\rm ind}$.  At energies much lower than $\Delta_{\rm ind}$, we can write a  low-energy effective Hamiltonian which involves the six MBSs
\begin{equation}
H_{\rm sys}=it_{12}\gamma_1'\gamma_{2}'+ it_{23}\gamma_2'\gamma_{3}'+it_{31}\gamma_3'\gamma_{1}'-i\sum_{j=1}^3 E_j\gamma_j\gamma_j'\,,
\label{h6maj}
\end{equation}  
where $\gamma_{j}'$ ($\gamma_{j}$) stand for the inner (outer) MBSs in wire $j=1,2,3$,  $t_{12},t_{23},t_{31}$ are the coupling between the inner MBSs and $E_j$ are the coupling strengths between the MBSs within each wire. For example, the braiding could be implemented by controlling the tunneling matrix elements by affecting some fluxes acting on the split Josephson junctions at the trijunction position.\cite{vanHeck2012}
Note that although these phenomelogical parameters depend on the microscopic details characterizing the band structure of the semiconducting wires and also on external parameters, the form of $H_{\rm sys}$ is general and universal.

Because the central MBS are  in close proximity, they overlap. Defining $E_M=\sqrt{t^2_{12}+t^2_{23}+t^2_{31}}$ the overlap energy scale, we can therefore consider the limit $E_M\gg E_j$ for long wires. Only one linear combination of the $\gamma_j'$ drops out from the Hamiltonian while the other two form a 2-dimensional subspace at energy  $\pm(1/2) E_M$. At energy much smaller than $E_M$, one can thus project $H_{\rm sys}$ in the left 4-dimensional space to obtain:\cite{vanHeck2012}
\begin{align}
H_{Y}=\frac{i}{2}\sum_{j=1}^3B_j\gamma_0\gamma_j\,,
\label{h4maj}
\end{align}
where $\gamma_0=(1/\sqrt{3})(\gamma_1'+\gamma_2'+\gamma_3')$, and $B_j=-4 E_j\frac{t_{jj+1}}{E_M}$.
We model the microwave cavity by a single mode electric field ${\bs {\hat E}}={\bs E}_0(a^\dagger+a)$, with ${\bs E}_0$ and $a$ ($a^\dagger$) being the vector amplitude of the electric field  and the annihilation (creation) operator for photonic field in the cavity, respectively. Taking into account the electric dipole matrix elements between the cavity electric field and the wires \cite{HasslerNJP2014,GinossarNatComm2014} together with  the capacitive coupling between the wires and underlying superconducting cavity \cite{trif2012,Dmytruk2015,Dartiailh2016}, we can write an effective coupling between the cavity electric field and the four MBSs. It amounts to substitute $B_j\rightarrow B_j[1+\alpha_j(a^\dagger+a)]$. The coeficients $\alpha_j$ depend on the detail microscopic Hamiltonians. We assumed the coupling between the MBSs and the photons to  take into account the exponential coupling of the MBSs as in Refs.~[\onlinecite{KnappPRX2016,SauCM2018,Trif2019}]. Physically, this means that far apart MBSs cannot be affected by the cavity field, by definition of being topologically protected.
The low-energy effective  Hamiltonian for the MBSs in the cavity thus reads in a compact form 
\begin{align}
H_{Y}(t)=\frac{i}{2}\gamma_0[{\bs B}(t)+{\bs g}(t)(a^\dagger+a)]\cdot{\bs \gamma}\,,
\end{align}
with ${\bs B}$,  ${\bs g}$ and ${\bs \gamma}$ are 3-dimensional vectors with $g_j=B_j\alpha_j$. Because  ${\bs g}\nparallel{\bs B}$ there is a dynamics triggered by the photonic field on the Majorana states when ${\bs B}(t)$ depends on time, which is the case when the MBSs are braided.

In order to make progress with this Hamiltonian, we introduce usual fermionic operators written in terms of the MBSs ones as $c_1=(\gamma_1-i\gamma_2)/2$ and $c_2=(\gamma_0-i\gamma_3)/2$. This allows us to rewrite the Hamiltonian $H_{Y}$ in the basis $\{|00\rangle, c_1^\dagger|00\rangle=|10\rangle,c_2^\dagger|01\rangle=|00\rangle,c_1^\dagger c_2^\dagger|00\rangle=|11\rangle\}$. In that purpose, it is first worth noticing that
the pair of states  $\{|00\rangle,|11\rangle\}$ and $\{|01\rangle,|10\rangle\}$ do not couple with each other as a consequence of the parity conservation in the system. It is therefore convenient to introduce the
parity operator defined by $\tau_z=\gamma_0\gamma_1\gamma_2\gamma_3$ which has two eigenvalues  $\tau=\pm$.
For a given parity, $H_{Y}$ takes the following very simple form (after a $-\pi/2$ rotation around the $z-$axis on the unit sphere):
\begin{equation}
H_{\rm Y}^\tau(t)=\frac{1}{2}{\bs B}_{\tau}^{\rm tot}(t)\cdot{\bs \sigma}+\omega_0 a^\dagger a~,
\label{Rabi}
\end{equation} 
with ${\bs B}_{\tau}^{\rm tot}(t)=[\tau B_{1}^{\rm tot}(t),B_{2}^{\rm tot}(t),B_{3}^{\rm tot}(t)]$ and ${\bs B}^{\rm tot}(t)={\bs B}(t)+{\bs g}(t)(a^\dagger+a)$. Due to the conservation of parity, the Hamiltonian $H_{\rm Y}^\tau(t)$ acts in a 2-dimensional space which is embodied by the Pauli matrices $\sigma_{x,y,z}$. 

The Hamiltonian in Eq. \eqref{Rabi} thus describes an effective spin $1/2$ coupled to a cavity mode similar to the Rabi model.  Thanks to this mapping, we can write the time-dependent evolution of the full density matrix of the system during a braiding operation. This is what we do in the next section.

\section{Dynamics of the system} 
\label{sec2}

By changing, in a time-dependent fashion,  the parameters of the above Hamiltonian, one can effectively implement the braiding of two chosen (external) MBSs in the trijunction (the middle one acts as an ancilla MBS). The goal here is to analyze such dynamics in the presence of the cavity field monitoring the braiding, as well as in the presence of various relaxation channels. We  address this issue in two fashions. First, we focus on the adiabatic limit, during the course of one adiabatic cycle and investigate that within the adiabatic master equation pertaining to the Majorana braiding. Second,  we address the dynamics  from a Floquet description, when the  braiding is performed continuously in a time-periodic fashion.  We will then compare the two approaches. 

\subsection{Adiabatic Master Equation}  

In the following, we assume that all manipulations are adiabatic with respect to the transitions 
between the low-energy Hilbert space and the first excited states, i.e.
that there are no real excitations outside the degenerate subspace.  The system Hamiltonian can be supplemented by the coupling to the environment. Both the Majoranas and the photons are assumed to be coupled to their own environment. For example, the Majoranas are subject to the phonons in the material, while the cavity photons are coupled to the external photons in the transmission lines. The full Hamiltonian can be written as $H_{\rm tot}^\tau=H_{\rm Y}^\tau(t)+H_{e}+H_{int}$, with the second and third terms being  the Hamiltonian of the environment and its coupling to the system, respectively. The dynamics of the total system density matrix is governed by the Liouville equation $\dot{\rho}_{\rm tot}^\tau=i[H_{\rm tot}^\tau, \rho_{\rm tot}]$.  The reduced density matrix for the system is obtained by partially tracing over the environmental degrees of freedom,  $\rho_\tau(t)={\rm Tr}_e[\rho_{\rm tot}^\tau(t)]$. We consider weak coupling to the environment and apply the standard Born and Markov approximations in the interaction picture, and subsequently trace over the bath degrees of freedom.  Using the secular approximation and assuming that the environment does not change the parity of the system,  we can express the master equation as follows (we set $\hbar=1$ hereafter):
\begin{align}
&\dot{\rho}_\tau(t)=-i[H_{\rm Y, eff}^\tau(t), \rho_\tau(t)]+\\
&\frac{1}{2}\sum_{n}\Gamma_{nm}(t)\left(2P_{\tau,nm}\rho_\tau P_{\tau,nm}^\dagger-\{P_{\tau,nm}^\dagger P_{\tau,nm},\rho_\tau\}\right)\nn
\end{align}
with
\begin{align}\label{defPnm}
P_{\tau,nm}=\left\{
\begin{array}{cc}
|\psi_n^\tau(t)\rangle\langle\psi_m^\tau(t)| &\,\, {\rm for}\,\Delta E_{nm}>0\\
|\psi_m^\tau(t)\rangle\langle\psi_n^\tau(t)| &\,\, {\rm for}\, \Delta E_{nm}<0,
\end{array}
\right.
\end{align} 
and where $\{\cdot,\cdot\}$ denotes the anticommutator.
We also introduced  
\begin{align}
H_{\rm Y, eff}^\tau(t)&=\sum_{n,m}[E_{n}(t)\delta_{n,m}-A_{nm}^{\tau}(t)]\,|\psi_n^\tau(t)\rangle\langle\psi_m^\tau(t)|,
\end{align}
where $E_{n}(t)$ and $|\psi_{n}^\tau(t)\rangle$ are the instantaneous eigenvalues (independent of $\tau$) and eigenstates (dependent on $\tau$) of the Majorana-photon Hamiltonian,  while  $A_{nm}^\tau(t)=i\langle\psi_n^\tau(t)|\dot{\psi}_m^\tau(t)\rangle$ represents the gauge field associated with the dynamics.  Note that the above expression is in the standard  Lindblad form and therefore, at this order of the dynamics, the positivity of the density matrix is conserved  during the time evolution. The rates $\Gamma_{nm}(t)$ are found from the Fermi's golden rule applied for the instantaneous eigenstates and eigenvalues:
\begin{equation}
\Gamma_{mn}(t)=2\pi|\langle\psi_n^\tau(t)|H_{int}|\psi_m^\tau(t)\rangle|^2S_{e}[E_{n}(t)-E_{m}(t)]\,,
\end{equation}  
with $S_{e}[\omega]$ the spectral function of the equilibrium fluctuations in the environment (that depends on the particular bath considered). 
We focus on the stationary regime  and thus require $\dot{\rho}_\tau=0$. In leading order in the dynamics, the diagonal terms follow the usual equilibrium distribution found from the detailed balance equation, and  are independent of the parity $\tau$, i.e. $\rho_{\tau,nm}\equiv\rho_{nn}$. The off-diagonal terms instead depend solely on the dynamics and read: 
\begin{align}
\rho_{\tau,nm}&=\frac{A_{nm}^{\tau}(t)(\rho_{nn}-\rho_{mm})}{(E_{n}-E_{m})+\frac{i}{2}\sum_{n'}[\Gamma_{mn'}(t)+\Gamma_{n'n}(t)]}\nn\\
&\approx \frac{A_{nm}^{\tau}(t)(\rho_{nn}-\rho_{mm})}{E_{n}-E_{m}}\,.
\end{align}
and $\Gamma_{nm}=\Gamma_{mn}\exp{[\beta(E_{n}-E_m)]}$. The expectation value of an observable $O$ thus reads
\begin{align}
\langle O_\tau(t)\rangle\approx O^{0}(t)+\sum_{n\neq m}\frac{O_{mn}A_{nm}^{\tau}(t)(\rho_{nn}-\rho_{mm})}{E_{n}-E_{m}}\,,
\end{align}
with $O^{0}(t)=\sum_n\langle\psi_{n,\tau}|O|\psi_{n,\tau}\rangle\rho_{nn}$ being the instantaneous expectation value of the operator $O$. Let us call $\langle \Delta O_\tau(t)\rangle\equiv\langle O_\tau(t)\rangle-O^{0}(t)$ the deviation from the ``static'' expectation value, pertaining only to the dynamics. We stress here that the density matrix itself can depend on the dynamics, and thus Berry phase contributions can occur in the diagonal terms too. However, the effects are of second order in dynamics, and in leading  order we can safely assume the equilibrium distributions.

In order to find the evolution of a given operator $O$, we thus need to investigate the  instantaneous  Hamiltonian and associated eigenstates.  In the following, we consider the dispersive limit $|{\bs g}(t)|\ll|\omega_0-B(t)|\ll\omega_0,B(t)$, and we can thus resort to the rotating wave approximation (RWA) as well as perturbation theory in the spin-photon coupling. That is best investigated by performing a unitary transformation $U(t)$ that diagonalizes the bare spin Hamiltonian, and by keeping in the resulting terms only the RWA contributions. This means $H_{\rm Y}^\tau(t)\rightarrow \tilde{H}_{\rm Y}^\tau(t)=U_\tau(t)H_{\rm Y}^\tau(t)U^\dagger_\tau(t)$ so that
\begin{align}
\tilde{H}_{\rm Y}^\tau(t)=B(t)\sigma_z+\omega_0a^\dagger a+[\tilde{g}_\tau(t)\sigma_-a^\dagger+\tilde{g}_\tau^*(t)\sigma_+a]\,,
\end{align}   
with $\tilde{g}_\tau(t)=\tilde{g}_{x}(t)+i\tau\tilde{g}_{y}(t)$, $\tilde{{\bs g}}_\tau(t)\cdot{\bs \sigma}=U_\tau(t)({\bs g}_\tau(t)\cdot{\bs \sigma})U^\dagger_\tau(t)$, and $U_\tau(t)=(|\psi_{+}^{\tau}(t)\rangle,|\psi_{-}^{\tau}(t)\rangle)$ is a unitary matrix built from the instantaneous spin eigenstates diagonalizing the bare Hamiltonian.
They satisfy  ${\bs B}_\tau(t)\cdot{\bs\sigma}|\psi_{\sigma}^{\tau}(t)\rangle=\sigma B(t)|\psi_{\sigma}^{\tau}(t)\rangle$. We  write $|\psi_\sigma^{\tau}(t)\rangle=U^\dagger_\tau(t)|\sigma\rangle$ to simplify notations.  

With this description at hand, we can now treat  the resulting spin-photon coupling in perturbation theory. Note that we can also write $\tilde{g}_\tau(t)=|\tilde{g}(t)|\exp{[i\tau\Phi(t)]}$, with $|\tilde{g}(t)|=\sqrt{\tilde{g}_x^2(t)+\tilde{g}_y^2(t)}$ and $\Phi(t)=\arctan{[\tilde{g}_y(t)/\tilde{g}_x(t)]}$, which implies the parity only enters in the phase.   To  further diagonalize this interaction, we perform yet another unitary transformation $\tilde{U}_\tau(t)=e^{-S_\tau(t)}=1-S_\tau(t)+\dots$ which, up to leading order in $\tilde{g}_\tau(t)$, diagonalizes the full system Hamiltonian. 
Using $S_\tau(t)=\tilde{g}_\tau(t)/[\omega_0-B(t)]\sigma_-a^\dagger-{\rm h.c.}$,
we find  
\begin{align}
\tilde{H}_{\rm Y}^\tau(t)&=\left[B(t)+\frac{|\tilde{g}(t)|^2}{\omega_0-B(t)}(a^\dagger a+1/2)\right]\sigma_z\nn\\
&+\omega_0a^\dagger a+\mathcal{O}(\tilde{g}_\tau^3(t))\,,
\end{align}
for the  resulting diagonal Hamiltonian and
\begin{align}
|\psi_{n}^\tau(t)\rangle&=U_\tau^\dagger(t)e^{S_\tau(t)}(|\sigma\rangle\otimes|n\rangle)\,,
\end{align}
for the eigenstates.  We stress that $|\tilde{g}_\tau(t)|\equiv|\tilde{g}(t)|$ is independent of $\tau$ and thus the instantaneous Hamiltonian does not reveal the parity. We now evaluate the gauge field contribution to the effective Hamiltonian and find, up to second order in the coupling to the cavity $A_{nm}^\tau(t)=A_{\tau, nm}^{(0)}(t)+A_{nm}^{(1)}(t)+A_{nm}^{(2)}(t)$, with:
\begin{align}
A_{\tau}^{(0)}(t)&=iU^\dagger_\tau(t)\dot{U}_\tau(t)\approx \tau(1-\sigma_z\cos{\theta})\dot{\phi}\,,\\
A_{\tau}^{(1)}(t)&=i\dot{S}_\tau(t)-[S_\tau,A_{\tau}^{(0)}(t)]\nn\\
&\approx\frac{\tilde{g}_\tau(t)}{\omega_0-B}\left[i\frac{|\dot{\tilde{g}}(t)|}{|\tilde{g}(t)|}+\tau\Phi_{t}^{B}\right]a^\dagger\sigma_-+{\rm h. c. }\,,\\
A_{\tau}^{(2)}(t)&=[S_\tau,[S_\tau, A_{\tau}^{(0)}(t)]]+i\frac{S_\tau(t)\dot{S}_\tau-\dot{S}_\tau S_\tau(t)}{2}\nn\\
&\approx 2\tau\Phi_{t}^{B}\frac{|\tilde{g}(t)|^2}{(\omega_0-B)^2}(a^\dagger a+1/2)\sigma_z\,,
\end{align}
where $\Phi_{t}^{B}\equiv (\dot{\Phi}-\cos{\theta}\dot{\phi})$ is the effective angular velocity. We only keep  above the terms that  shift the instantaneous eigenvalues, as well as the off-diagonal terms that correspond to transitions involving energies $\pm|\omega_0-B|$ (a spin flip with emission/absorbtion of a cavity photon). All the other terms are assumed small and thus negligible in the adiabatic limit considered here.  For simplicity, we  assumed that $B(t)\equiv B$ (constant instantaneous splitting).

Next we use the above findings to evaluate various photonic observables, such as the average photon number $\langle n(t)\rangle$ and the second order coherence function at zero delay time $g^{(2)}(0)$. The latter is given as \cite{Milburn}:
\begin{equation}
    g^{(2)}(0)=1+\frac{V_n-\langle n\rangle}{\langle n\rangle^2}\,,
\end{equation}
with  $V_n=\langle n^2\rangle-\langle n\rangle^2$ being the variance of the photon number in the cavity, and it quantifies the statistics of the photons emitted into the cavity. The photons  can then be accessed in photon transmission or reflection measurements. We find for
the average photon number in the cavity:
\begin{align} \label{eq:nt}
   \!\!\! \langle n(t)\rangle&=\langle n\rangle_0-\frac{|\tilde{g}(t)|^2}{(\omega_0-B)^2}\left(1+\tau\frac{\Phi_t^B}{\omega_0-B}\right)F\,,\nonumber\\
\end{align}
where we defined $\langle n\rangle_0=\sum_{k,\sigma}\rho_{n,\sigma}n$ as being the average number of photons in the absence of the coupling to the MBSs, and 
\beq \label{eq:defF}
F=\sum\limits_k (\rho_{k+1\downarrow}-\rho_{k\uparrow})(k+1)\,,
\eeq
 is a function that can depend on temperature. The second and third terms in \eqref{eq:nt} account for the instantaneous and Berry phase contribution to the photon number, respectively. 
This expression fulfills the right limits.  For example at zero temperature, the system density matrix is $\rho_{k\sigma}=\delta_{k,0}\delta_{\sigma,\downarrow}$, and the photons are totally decoupled from the Majorana dynamics. In order for the photonic state to get imprinted with the geometry of the spin trajectory the excited states (either spin or photon) need to be thus populated.  In particular, let us focus on the contribution that depends on the parity in the above expression (labelled as  $\langle\Delta n_\tau(t)\rangle$ from here on). The average number of photons due to the braiding dynamics can be found by integrating over the time the  braiding is performed,  i.e.  $\langle \Delta n_\tau\rangle=(1/T)\int_0^T dt\langle \Delta n_\tau(t)\rangle$. Assuming, for simplicity, that the braiding is done at constant velocity,  the average photon number in the cavity reads:
\begin{equation}\label{eq:nph1}
\langle\Delta n_\tau\rangle=\tau\frac{(2\pi N-2\phi_B)}{T}\frac{|\tilde{g}|^2_{av}}{(\omega_0-B)^3}F\,,
\end{equation}
with $N$ quantifying the number of times the phase $\Phi$ winds during the cycle,
$\phi_B$ denotes the geometric Berry phase on the spin Bloch sphere which, for our specific case is $\phi_B=\pi/4$.   Also, $|\tilde{g}|^2_{av}=(1/T)\int_{0}^Tdt|\tilde{g}(t)|^2$.  

Next we analyze the photon fluctuations, and in particular we evaluate the quantity $V_n'(t)\equiv V_n(t)-\langle n(t)\rangle$ that enters the expression for $g^{(2)}(0)$. Let us now focus  only on the contribution stemming from the dynamics, labelled $\Delta V'_{n,\tau}(t)$, we get:
\begin{align}
\Delta V'_{n,\tau}(t)&=-2\tau\frac{\Phi_t^B|\tilde{g}(t)|^2}{(\omega_0-B)^3}G\,,
\end{align}
with 
\beq\label{eq:defG}
G=\sum\limits_k(\rho_{k+1\downarrow}-\rho_{k\uparrow})(k+1)(k-\langle n_0\rangle)\,.\eeq
 That in turn allows us to evaluate the contribution of the dynamics to the second order photon coherence function, written as $g^{(2)}(0)=g^{(2)}_i(0)+g^{(2)}_{\tau}(0)$, where the first and last term  stand for the instantaneous and non-adiabatic contributions, respectively. For the latter, we obtain:
\begin{align} 
g^{(2)}_{\tau}(0)&=-2\tau\frac{\Phi_t^B|\tilde{g}(t)|^2}{\langle n\rangle_0^2(\omega_0-B)^3}\left[G+\langle n\rangle_0(g^{(2)}_0-1)F\right]\,,
\label{eq:Vph1}
\end{align}
where $g_0^{(2)}$ is the photon second order coherence function in the absence of the coupling to the MBSs.  Note that we can again evaluate the average of this quantity over one braiding period, in which case the dependence on the Berry phase is exactly the same as that found above for the photon number. We thus see that the statistics of the photons too is altered by the geometry and the parity of the MBSs trajectory during the braiding. While such effects are weak in the dispersive regime, they are expected to be more pronounced close to the resonance condition $\omega_0\sim B$. Such regime is beyond the scope of the present paper, but it is an interesting question in the future. 


Finally, let us also  evaluate  the spin expectation value, in particular  along the instantaneous magnetic field direction. In leading order in the dynamics, we get 
\begin{equation}
\langle\Delta\sigma^{z}_{\tau}(t)\rangle=-\langle\Delta n_\tau(t)\rangle\,,
\label{eq:sp1}
\end{equation}
which is a consequence of the conservation of the $C_z=a^\dagger a+\sigma_z$ within the RWA, while the transverse components vanish at this order. 

Eqs. (\ref{eq:nph1}),  (\ref{eq:Vph1}), and (\ref{eq:sp1}) that show that the average number of photons, the second order coherence function, and the spin expectation value  are in direct correspondence with the braiding Berry phase $\phi_B$ and the parity $\tau$ in the adiabatic regime represent the main results of this section. 

In order to probe the MBSs and their dynamics, one can access the cavity itself and record the response. We thus add an extra term into the total Hamiltonian that accounts for the cavity driving, $H_d=\epsilon a^{-i\omega_dt}+{\rm h. c. }$, with $\epsilon$ and $\omega_d$ being the amplitude and the frequency of the driving field. Such driving can be performed by sending a coherent states onto the cavity through the input ports.  That results in adding an extra term 
$\Delta\dot{\rho}_{nm}=i\langle\psi_n^\tau|[H_d(t),\rho]|\psi_m^\tau\rangle$,
in the density matrix evolution and then calculate the response. We leave such a discussion for a future study.

Let us now  qualitatively explain why the microwave photons are sensitive to the Berry phase accumulated by the effective spin during a braiding protocol. The coupling between the photons and the Majorana zero modes is encapsulated in the term ${\bs g}\cdot{\bs \sigma}(a^\dagger+a)$. When the MBSs are braided, this amounts the effective spin to follow a trajectory on the Bloch unit sphere which corresponds to an octant.\cite{KarzigPRX2016} Such trajectory in parameter space directly imprints the photon field  via non-adiabatic corrections.  In every braiding
operation, thus in every close octant trajetory, the system acquires an extra  Berry phase $\phi_B=\pi/4$. If the braiding operation is repeatedly done at some driving frequency $\Omega$,  the extra accumulated phase over a time period $T=1/\Omega$ is $\phi_B$ which has 
the effect of a phase shift $\exp(i \phi_B \Omega T)$ in the dynamics. Equivalently, such Berry phase shift 
shows up as a frequency  or energy shift $\propto\Omega\phi_B$. The
Berry phase thus becomes thus measurable spectroscopically (i.e. by a frequency shift which can be detected by standard dispersive readout techniques) by repeating the braiding
operation many times with a frequency $\Omega=1/T$.\cite{Trif2019} We can actually provide more substance to this picture by developing a Floquet description of the dynamics. This is what we do next.

\subsection{Floquet description of the dynamics}

Here we briefly describe the braiding as a periodic process and make use of the  Floquet formalism. That will also allow us to calculate, among other things, the stationary number of photons in the cavity and the resulting photon statistics.  The trijunction Hamiltonian satisfies $H_{\rm Y}^\tau(t+T)=H_{\rm Y}^\tau(T)$, and thus the general solution of the time-dependent Schrodinger equation reads $|\psi_n^\tau(t)\rangle=e^{-i\epsilon_{\tau,n}t}|\Psi_n^\tau(t)\rangle$, with $\epsilon_{\tau,n}$ being the quasienergy of the Floquet state $n$ with parity $\tau$ (as opposed to the adiabatic case, here $\epsilon_{\tau,n}$ depends on parity as it contains intrinsically the Berry phase) and $|\Psi_n^\tau(t+T)\rangle=|\Psi_{n}^\tau(t)\rangle$ being the corresponding (periodic) Floquet state.  The evolution operator can also be readily expressed in terms of the same quantities as $U_{Y,\tau}(t)=\sum_ne^{-i\epsilon_{n,\tau}t}|\Psi_n^\tau(t)\rangle\langle\Psi_n^\tau(0)|$. 
Next we define 
$\Delta_{nm}^\tau(k)=\epsilon_{\tau,n}-\epsilon_{\tau,m}+\tau k\Omega$ and  follow closely the derivations in  Ref.~[\onlinecite{BlumelPRA1991}]. Using the  Floquet description and within the Markov and secular approximations, we can write the evolution of the system density matrix (in the interaction picture) as follows:
\begin{align}
\dot{\rho}_\tau(t)=\frac{1}{2}\sum_{nmk}&\Gamma_{nm}^\tau(k)\left(2P_{\tau,nm}\rho_\tau(t) P_{\tau,nm}^\dagger \right.\nn\\
&-\left.\{P_{\tau,nm}^\dagger P_{\tau,nm},\rho_\tau(t)\}\right)
\end{align}
with $P_{\tau,nm}$ defined in Eq. \eqref{defPnm}
 and 
\begin{align}
\Gamma_{nm}^\tau(k)=2\pi|H_{int}^{\tau,nm}(k)|^2S_{e}[\Delta_{nm}^\tau(k)]\,,
\end{align}  
where $H_{int}^{\tau,nm}(k)=\int_{-\infty}^\infty dte^{-ik\Omega t}\langle\Psi_n^\tau(t)|H_{int}|\Psi_m^\tau(t)\rangle$. It is convenient to express the matrix elements of the density matrix in the Floquet basis, $\rho_{\tau,nm}(t)=\langle\Psi_n^\tau(0)|\rho(t)|\Psi_m^\tau(0)\rangle$, which in the stationary limit obeys:
\begin{align}
\rho_{\tau,nn}(t)\sum_m\Gamma_{nm}^\tau&-\sum_{m}\rho_{\tau,mm}(t)\Gamma^\tau_{nm}=0\,,\\
\Gamma_{nm}^\tau&=\sum_{k=-\infty}^\infty\Gamma^\tau_{nm}(k)\,,
\end{align} 
and $\rho_{\tau,nm}(t)=0$, for $n\neq m$, i.e. the off-diagonal matrix elements of the density matrix in the Floquet basis are zero. We mention here that the $k=0$ terms in the above expressions quantify the static contributions, in the absence of the driving of the MBSs, while all $k\neq0$ amount for exchange of a $k\Omega$ energy exchange  with the driving field. The diagonal part of the density matrix can be found explicitly from the above Floquet rate equation and for a specific environment. We will not proceed with such a calculation here, but just mention in passing that since it depends on the Floquet states, it eventually contains both the parity and the Berry phases of the Majorana trajectory. Nevertheless, in the case of slow driving, they correspond to the equilibrium thermal distribution. An observable $O$ (such as the number of photons in the cavity and the spin expectation value) can be evaluated in the stationary Floquet state as $\langle O(t)\rangle=\sum_{p}\langle \Psi_p^\tau(t)|O|\Psi_p^\tau(t)\rangle\rho_{pp}^\tau$. In order to find the Floquet states of the combined system one needs to proceed numerically. However, if  we restrict now to the dispersive regime, in which case the cavity is detuned from the Floquet energies associated with the Majorana dynamics, we can treat the problem perturbatively.  For that, let us express the Schrodinger equation as $\mathcal{H}_{\rm Y}^\tau(t)|\Psi_n^\tau(t)\rangle=\epsilon_{\tau,n}|\Psi_n^\tau(t)\rangle$, with $\mathcal{H}_{\rm Y}^\tau(t)=H_{\rm Y}^\tau(t)-i\partial/\partial t$, and define the unitary transformation $\tilde{U}_\tau(t)=e^{-S_\tau(t)}$, with $S_\tau(t)$ chosen so that it diagonalizes the   $\mathcal{H}_{\rm Y}^\tau(t)$ in leading order in the Majorana-photon coupling.  Keeping only the rotating terms (therefore using a generalized RWA for the Floquet case), we obtain:
\begin{align}
S_\tau(t)\approx\frac{1}{2}\sum_{n\neq m,q} &[1+{\rm sign}(\Delta^\tau_{nm}(q))]\nn\\
&\times\frac{e^{iq\Omega t}g_{\tau,nm}^{q}}{\omega_0-\Delta^\tau_{nm}(q)}\Sigma^\tau_{nm}(t)a^\dagger-{\rm h.c.}\,,
\end{align}
with  $\Sigma^\tau_{nm}(t)\equiv|\Psi_{0,n}^\tau(t)\rangle\langle\Psi_{0,m}^\tau(t)|$ and $g_{\tau,nm}^{q}=\int dte^{-iq\Omega t}\langle\Psi_{0,n}(t)|{\bs g}(t)\cdot{\bs \sigma}|\Psi_{0,m}(t)\rangle$, where  $|\Psi_{0,n}^\tau(t)\rangle$ is  the $n-$th bare Floquet state in the absence of the coupling to the cavity. Also, ${\rm sign}(\Delta^\tau_{nm}(q))$ selects only the rotating terms in the expression for $S_\tau(t)$.     Note that for  $k=q=0$ the transformation $S_\tau(t)\equiv S_\tau$ reduces to that found in the static (adiabatic) case. To summarize the effects of the above transformation, we found that a mixed Majorana-photon Floquet state $|\Psi_{n}^\tau(t)\rangle$ can be written as:
\begin{align}
|\Psi_{\tilde{n}}^\tau(t)\rangle&=e^{S_\tau(t)}|\Psi_{0,n}^\tau(t)\rangle\otimes|m\rangle\nn\\
&\approx[1+S_\tau(t)+S^2_\tau(t)/2]|\Psi_{0,n}^\tau(t)\rangle\otimes|m\rangle\,,
\end{align}
where on the right side $|m\rangle$ quantifies the photonic state with $m$ photons, 
while $\tilde{n}$ labels the final Majorana-photon Floquet state.  We now assume slow dynamics and low temperature compared to the level splittings, so that $\rho_{\tau,pp}$ follows the static thermal distribution, and it is thus independent on the parity $\tau$.  As examples, let us  calculate  again the average value of the number of photons in the cavity $\langle n(t)\rangle\equiv\langle a^\dagger a\rangle$ and the variance $V'_{n}(t)=\langle n^2(t)\rangle-\langle n(t)\rangle^2-\langle n(t)\rangle$, focusing only on the contributions due to the coupling to the dynamical MBSs (disregarding the bare photon number/bare variance).  For that, let us assume $\Delta^\tau_{+-}(q)\equiv \Delta^\tau_0(q)>0$ and write $\rho_{nn}\equiv\rho_{n,\sigma}$, with $n$ and $\sigma$ quantifying the photons and the Floquet index, respectively.  
We find
\begin{align}
&\langle n_\tau(t)\rangle
&=\sum_{q,q'}\frac{{\rm Re}[e^{2i(q-q')\Omega t}g_{\tau,\uparrow\downarrow}^{q}g_{\tau,\downarrow\uparrow}^{-q'}]}{[\omega_0-\Delta^\tau_{0}(q)][\omega_0-\Delta^\tau_{0}(q')]}F\,,
\end{align}
for the Majorana-induced change in the photon number where $F$ has been defined in Eq. \eqref{eq:defF}. For the deviation of the  variance $V_{n,\tau}'(t)$, we obtain:
\begin{align}
V'_{n,\tau}(t)&=2\sum_{q,q'}\frac{{\rm Re}[e^{2i(q-q')\Omega t}g_{\tau,\uparrow\downarrow}^{q}g_{\tau,\downarrow\uparrow}^{-q'}]}{[\omega_0-\Delta^\tau_{0}(q)][\omega_0-\Delta^\tau_{0}(q')]}G\,.
\end{align}
where $G$ has been defined in Eq. \eqref{eq:defG}.

We mention that the Floquet quasi-energy can be written as $\epsilon_{\tau,n}=\bar{\epsilon}_{\tau,n}-\phi_{n,G}^\tau/T$, where $\bar{\epsilon}_{\tau,n}=(1/T)\int_0^Tdt\langle\Psi_{0,n}^\tau(t)|H_Y(t)|\Psi_{0,n}^\tau\rangle$ and $\phi_{n,G}^\tau=i\int_0^Tdt\langle\Psi_{0,n}^\tau(t)|\partial_t|\Psi_{0,n}^\tau\rangle$ are the average Floquet energy and the  geometrical (non adiabatic in general) phase, respectively. In the adiabatic limit, $\bar{\epsilon}_{\tau,n}$ becomes the good instantaneous eigenstate of the system, while $\phi_{n,G}^\tau\rightarrow\phi_{n,B}^\tau$, i.e. the Berry phase associated with the adiabatic motion. 
Once again, we can evaluate the average number of photons in the cavity over one period, $\langle n_\tau\rangle=(1/T)\int_0^t dt\langle n_\tau(t)\rangle$, and retaining only the terms that are linear in $\Omega$ (geometric in origin) we obtain:
\begin{align}\label{eq:nph2}
\langle n_\tau\rangle&=\tau\sum_q\frac{(2\pi q-2\phi_B)}{T}\frac{|g_{\uparrow\downarrow}^{q}|^2}{(\omega_0-B)^3}F\,,
\end{align}
where $\phi_B$ is the braiding geometric phase.
This expression has the same form as the one found in Eq. (\ref{eq:nph1}).
 The first term accounts for the winding of the phase of the complex Majorana-photon coupling, while the second one pertains to the bare Berry phase associated with the Majorana braiding. The very same conclusions apply to $g^{(2)}(0)$, which affect the statistics of the photons. Note, however, that in the Floquet description resonances can drastically enhance the number of photons in the cavity which be strongly be affected by the Berry phases, as they move the levels on or off resonance. Such a regime is beyond the scope of our work. 
  
\section{Discussion }
\label{discussion}

\subsection{Majorana versus Andreev bound states}

Since the first transport experiments on topological semiconducting wires which found signatures of Majorana bound states in transport quantities, they have been a nagging question whether the observed robust zero-biased peaks might as well  be interpreted as near-zero energy  Andreev bound states (ABSs). It is thus important to address the question whether the dynamics can differentiate between MBSs and near-zero ABSs during a braiding scheme,  as well as its detection within our cavity QED scheme. We thus alter our trijunction by assuming one of the wires (the wire 3) is hosting a local near-zero ABS at the junction, instead of two separated Majorana modes at its two extremities. The goal is then to exchange the two external  MBSs hosted in the other two wires following the same braiding steps as before. The ABS  is assumed to be split by an energy $U_3$ which we may suppose  is subject to the same degree of control as for the topological regions. The two MBSs at the junction in the topological wires will couple equally to the ABS. In the limit of strong tunneling at the junction, and assuming zero splittings for the MBSs, as well as for the ABS, the low-energy 
Hamiltonian now reads:
\begin{equation}
H_{Y}^A=U_3(2c_A^\dagger c_A-1)+i\sum_{j=1,2}t_{j3}\gamma_j'(e^{i\phi_{j}}c_A^\dagger+{\rm h. c.})\,,
\end{equation}
where $c_A$ ($c_A^\dagger$) are the annihilation (creation) operator for the Andreev state, $t_{j3}$ the coupling between the inner MBSs and the ABS, while $\phi_{1,2}$ are angles that quantify the phase of the couplings ($\phi_1\neq\phi_2$ in general). Physically, $e^{i\phi_{j}}c_A^\dagger+{\rm h. c.}\equiv\gamma_3'$ defines a Majorana fermion (the partner  Majorana is $\gamma_3=i(e^{i\phi_{j}}c_A^\dagger-{\rm h. c.})$). For $\phi_1=\phi_2$ the situation reduces to the previous case. However, generally $\phi_1\neq\phi_2$, and the coupling between the inner MBSs $\gamma_{1,2}'$ and the ABS is not symmetric anymore which, as we show in the following, affects the braiding. We assume a small asymmetry, or $\delta\phi=\phi_2-\phi_1\ll1$, and treat the resulting contribution in perturbation theory along with the intra-wires splittings. The resulting $4$-MBSs' Hamiltonian then has the same form as in Eq.~\eqref{h4maj}, with the only difference being that $B_3'=B_3+\delta B_3$, with $\delta B_3=\delta\phi\,t_{13}^2/E_M$. The last contribution stems from the anisotropic coupling to the ABS and vanishes for the case of separated Majoranas. To keep the discussion simple, we assume that during the braiding steps this term remains constant, in which case the Berry phase accumulated by the ground state becomes:
\begin{equation}
    \gamma_{\tau,B}^A=\tau(1-\cos{\Theta})\frac{\pi}{4}\,,
\end{equation}
with $\cos{\Theta}=\delta B_3/\sqrt{\Delta_1^2+\Delta_2^2+\delta B_3^2}$. We note that the spectrum is at all times degenerate in this picture, and thus no conclusion can be drawn from a spectral analysis. However, the above geometrical phase will be imprinted into the photonic field through the mechanism discussed before, and it is no longer universal for the ABS case. That could serve to differentiate between the two situations, Majorana vs. Andreev: in the former, the Berry phase imprinted is always $\tau\pi/4$, while in the latter it is non universal and it can vary between different runs or experiments.   

Finally, let us mention the simpler and maybe more generic case which occurs when  all effectively couplings are non zero. This leads to a direct  coupling between the neighboring MBSs at the Y-junction tip. In this  case, the ground state degeneracy is split and the two states acquire different dynamical phases in addition to the path dependent Berry phase contribution. The cavity is able to measure and distinguish these dynamical phases which are also non universal.  

\subsection{Experimental considerations}

 In order to operate the braiding protocol, the low-energy Majorana states shall remain isolated in energy from quasi-particles states which are inside the proximity induced gap (we demand $E_M\gg E_J$). At an experimental level, one may for example  use  epitaxially grown InSb nanowires with epitaxial Al on two of its six facets (see. Ref. \onlinecite{Lutchyn2018} for a recent review and refs therein). For these wires, the  induced superconducting pairing gap $\Delta_{\rm ind}$ is about the superconducting gap, namely $\Delta_{\rm ind}\sim\Delta\sim O(0.2meV)$.
Estimating $E_M$ as a fraction of the gap,  $E_M\sim \kappa \Delta$ with $\kappa\ll 1$, this puts an upper limit on $E_J$. For a wire of typical length $L\sim 2\mu$m, if $E_J$ is controlled by the overlap of the  Majorana wave functions, we can estimate $E_J\sim O(0.01meV)$\cite{Dmytruk2015} for a not too large magnetic field (as $E_J$ increases with the magnetic field). Although the condition $E_M\gg E_J$ may still be satisfied, it might be desirable to have a better tuning over the energy scale $E_J$. Another mechanism based on exchange of Majorana fermions in a network of superconducting nanowires controlled by Coulomb interactions rather than tunneling has been proposed.\cite{vanHeck2012} This would offer an extra knob to control $E_J$ and eventually to make it smaller. In order to adiabatically operate the braiding, one demands $\Omega\ll E_J$ which implies $\Omega\lesssim O(100)$MHz. Note that this is still much faster than the estimated quasiparticle poisoning which has recently been measured over 10ms.\cite{Marcus2015,Lutchyn2018}

\section{Conclusion} 
\label{conclusion}
In conclusion, we studied theoretically the dynamics of Majorana fermions in a Y-junction geometry coupled to a cavity and in the presence of dissipation. We investigated the stationary density matrix both in the adiabatic regime for the dynamics, as well as the case of periodic driving within the Floquet formalism. We consistently showed, within the two approaches, that the geometry of the trajectory as well as the parity of the ground state affect various light observables, such as the photon number in the cavity and the second order coherence function, which can be extracted from experiments by standard dispersive readouts techniques. We also discussed the situation in which one of the wire is supporting a near-zero energy ABS at the junction instead of a Majorana bound state and found that the  Berry phase measured by the cavity is generically non-universal. Finally, we also provided some  estimation on the different energy scales coming into play which imposes constraints on the braiding frequency for experiments.
In future studies, it would be interesting to go beyond the mean photon number and also address the (full-)statistics of the photons emitted in the cavity. It would be also very interesting to address the dynamics of the Majorana-Rabi model  in the presence of  non-classical states of light in the cavity (e. g. squeezed states or Fock states). Another obvious direction to pursue is to go beyond the weak light-matter interaction and analyze whether it would be possible to control the braiding process with light.

\vskip 0.5cm
{\bf Acknowledgments} \\
The International Centre for Interfacing Magnetism and Superconductivity with Topological Matter project (MT) is carried out within the International Research Agendas program of the Foundation for Polish Science co-financed by
the European Union under the European Regional Development Fund.

\vskip 0.5cm
{\bf Acknowledgments}\\
The authors declare no conflict of interest

\vskip 0.5cm
{\bf Keywords}\\
Topological superconductivity, Majorana fermions, non-Abelian braiding, geometric phase, cavity-QED.


\begin{thebibliography}{99}
\bibitem{kitaev2001unpaired} A. Y. Kitaev, {\it Phys.-Usp.} {\bf 2001},  44, 131.
\bibitem{alicea2012new} J. Alicea, {\it Rep. Prog. Phys.} {\bf 2012}, 75, 076501. 
\bibitem{franz2015review} S. R. Elliott, M. Franz, {\it Rev. Mod. Phys.} {\bf 2015}, 87, 137.
\bibitem{Aguado2017} R. Aguado, {\it Riv. Nuovo Cimento Soc. Ital. Fis.} {\bf 2017}, 40, 523. 
\bibitem{mourik2012signatures} V. Mourik, K. Zuo, S. M. Frolov, S. Plissard, E. Bakkers,  L. Kouwenhoven, {\it Science} {\bf 2012}, 336, 1003.

\bibitem{xu2012} M. T. Deng, C. L. Yu, G. Y. Huang, M. Larsson, P. Caroff, H. Q. Xu, {\it Nano Letters} {\bf 2012}, 12, 6414.
\bibitem{Das2012} A. Das, Y. Ronen, Y. Most, Y. Oreg, M. Heiblum, H. Shtrikman, {\it Nat. Phys.} {\bf 2012}, 8, 887.
\bibitem{albrecht2016exponential} S. Albrecht, A. Higginbotham, M. Madsen, F. Kuemmeth, T. Jespersen, J. Nygard, P. Krogstrup,  C. Marcus, {\it Nature} {\bf 2016}, 531, 206.
\bibitem{Deng2016} M. T. Deng, S. Vaitiekenas, E. B. Hansen, J. Danon, M. Leijnse, K. Flensberg, J. Nygard, P. Krogstrup, C. M. Marcus,
{\it Science} {\bf 2016}, 354, 1557.
\bibitem{Kouwenhoven2018} H. Zhang,  C.-X. Liu, S. Gazibegovic, D. Xu, J. A. Logan, G. Wang, N. van Loo, J. D. S. Bommer, M. W. A. de Moor, D. Car, R. L. M. Op het Veld, P. J. van Veldhoven, S. Koelling, M. A. Verheijen, M. Pendharkar, D. J. Pennachio, B. Shojaei, J. S. Lee, C. J. Palmstroem, E. P. A. M. Bakkers, S. Das Sarma, L. P. Kouwenhoven, {\it Nature} {\bf 2018}, 556, 74.
\bibitem{Lutchyn2018} R. M. Lutchyn, E. P. A. M. Bakkers, L. P. Kouwenhoven, P. Krogstrup, C. M. Marcus, Y. Oreg, 
{\it Nat. Rev. Mater.} {\bf 2018}, 3, 52.
\bibitem{Yazdani2014} S. Nadj-Perge, I. K. Drozdov, J. Li, H. Chen, S. Jeon, J. Seo, A. H. MacDonald, B. A. Bernevig, A. Yazdani, 
{\it Science} {\bf 2014}, 346, 602.
\bibitem{Wang2018} D. Wang, L. Kong, P. Fan, H. Chen, S. Zhu, W. Liu, L. Cao, Y. Sun, S. Du, J. Schneeloch, R. Zhong, G. Gu,
L. Fu, H. Ding, H.-J. Gao, {\it Science} {\bf 2018}, 362, 333.
\bibitem{Yazdani2019} B. J\"ack, Y. Xie, J. Li, S. Jeon, B. A. Bernevig, A. Yazdani, {\it Science} {\bf 2019}, 364, 1255.
\bibitem{Kitaev2003} A. Y. Kitaev, {\it Ann. Phys.} {\bf 2003}, 303, 2.
\bibitem{Alicea2011} J. Alicea, Y. Oreg, G. Refael, F. von Oppen,  M. P. A. Fisher, {\it Nat. Phys.} {\bf 2011}, 7, 412.
\bibitem{Sau2011} J. D. Sau, D. J. Clarke,  S. Tewari, {\it Phys. Rev. B} {\bf 2011}, 84, 094505.
\bibitem{vanHeck2012} B. van Heck, A. R. Akhmerov, F. Hassler, M. Burello,  C. W. J. Beenakker, {\it New J. Phys.} {\bf 2012}, 14, 035019.
\bibitem{SarmaRMP2008} C. Nayak, S. H. Simon, A. Stern, M. Freedman, S. Das Sarma, {\it Rev. Mod. Phys.} {\bf 2008}, 80, 1083. 
\bibitem{dassarma-majarona-review} S. Das Sarma, M. Freedman, C. Nayak, {\it Npj Quantum Information} {\bf 2015}, 1, 15001.
\bibitem{aasen2016-qc} D. Aasen, M. Hell, R. V. Mishmash, A. Higginbotham, J. Danon, M. Leijnse, T. S. Jespersen, J. A. Folk, C. M. Marcus,
K. Flensberg, J. Alicea, {\it Phys. Rev. X} {\bf 2016}, 6, 031016.
\bibitem{KarzigPRX2016} T. Karzig, Y. Oreg, G. Refael, M. H. Freedman, {\it Phys. Rev. X} {\bf 2016}, 6, 031019.
\bibitem{Trif2019} M. Trif, P. Simon, {\it Phys. Rev. Lett.} {\bf 2019}, 122, 236803.
\bibitem{KnappPRX2016} C. Knapp, M. Zaletel, D. E. Liu, M. Cheng, P. Bonderson, C. Nayak, {\it Phys. Rev. X} {\bf 2016}, 6, 041003.
\bibitem{HasslerNJP2014} C. Ohm, F. Hassler, {\it New J. Phys.} {\bf 2014}, 16, 126803.
\bibitem{GinossarNatComm2014} E. Ginossar, E. Grosfeld, {\it Nat. Commun.} {\bf 2014}, 5, 4772.
\bibitem{trif2012} M. Trif, Y. Tserkovnyak, {\it Phys. Rev. Lett.} {\bf 2012}, 109, 257002.
\bibitem{Dmytruk2015} O. Dmytruk, M. Trif, P. Simon, {\it Phys. Rev. B} {\bf 2015}, 92, 245432.
\bibitem{Dartiailh2016} M. C. Dartiailh, T. Kontos, B. Doucot, A. Cottet, {\it Phys. Rev. Lett.} {\bf 2017}, 118, 126803.
\bibitem{SauCM2018} A. Nag, J. D. Sau, {\it arXiv:1808.09939} {\bf 2018},  https://arxiv.org/abs/1808.09939.
\bibitem{Milburn} D. F. Walls and G. J. Milburn, {\it Quantum Optics},  Springer-Verlag, Berlin Heidelberg, {\bf 1994}.
\bibitem{BlumelPRA1991} R. Bl\"umel, A. Buchleitner, R. Graham, L. Sirko, U. Smilansky,  H. Walther, {\it Phys. Rev. A} {\bf 1991}, 44, 4521.
\bibitem{Marcus2015} A. P. Higginbotham, S. M. Albrecht, G. Kiršanskas, W. Chang, F. Kuemmeth, P. Krogstrup, T. S. Jespersen, J. Nyg\aa rd, K. Flensberg, C. M. Marcus, {\it Nat. Phys.} {\bf 2015}, 11, 1017.
\end{thebibliography}

\end{document}